\documentclass[prl,twocolumn,superscriptaddress,showpacs]{revtex4}

\usepackage{graphicx,soul}
\usepackage{color}
\usepackage{amsmath}
\usepackage{amsfonts}
\usepackage{amssymb}
\usepackage{bm}

\begin{document}

\title{Topological Polaritons and Excitons in Garden Variety Systems}

\author{Charles-Edouard Bardyn*}
\affiliation{Institute for Quantum Information and Matter, Caltech, Pasadena, California 91125, USA}
\author{Torsten Karzig*}
\affiliation{Institute for Quantum Information and Matter, Caltech, Pasadena, California 91125, USA}
\author{Gil Refael}
\affiliation{Institute for Quantum Information and Matter, Caltech, Pasadena, California 91125, USA}
\author{Timothy C. H. Liew}
\affiliation{Division of Physics and Applied Physics, Nanyang Technological University 637371, Singapore}

\begin{abstract}
Topological polaritons (aka topolaritons) present a new frontier for topological behavior in solid-state systems. They combine light and matter, which allows to probe and manipulate them in a variety of ways. They can also be made strongly interacting, due to their excitonic component. So far, however, their realization was deemed rather challenging. Here we present a scheme which allows to realize topolaritons in garden variety zinc-blende quantum wells. Our proposal requires a moderate magnetic field and a potential landscape which can be implemented, e.g., via surface acoustic waves or patterning. We identify indirect excitons in double quantum wells as a particularly appealing alternative for topological states in exciton-based systems. Indirect excitons are robust and long lived (with lifetimes up to milliseconds), and, therefore, provide a flexible platform for the realization, probing, and utilization of topological coupled light-matter states.
\end{abstract}




\maketitle



Topological phases and states in quantum systems have yielded a wealth of exotic phenomena, with measurable signatures at edges and surfaces. In electronic topological insulators, what otherwise would seem a simple semiconductor may exhibit conducting states at its edges~\cite{Hasan2010,Qi2011}. More recently, similar physics began to emerge in photonic systems. Theory~\cite{Haldane2008,Wang2008,Hafezi2011,Fang2012} was followed by demonstrations of photonic topological behavior in microwave-range photonic crystals~\cite{Wang2009} and arrays of coupled optical resonators~\cite{Hafezi2013,Ningyan2013} and waveguides~\cite{Rechtsman2012}, culminating with the realization of chiral edge states protected from backscattering. The prospects of photons propagating in a single direction --- thus circumventing material imperfections --- gave rise to the field of topological photonics, and may yet revolutionize photonic circuitry~\cite{Lu2014}.

These exciting recent developments, however, are by and large due to linear optical effects, while applications in photonic circuitry often crucially require nonlinear optical properties. Hence the importance of the recently proposed  ``topological polaritons''~\cite{Karzig2014} --- topological superpositions of excitons and photons --- which combine topology and nonlinear properties via light-matter interactions. Excitons interact with one another and, when placed inside of an optical microcavity, hybridize into so-called exciton-polaritons which balance a strong exciton nonlinearity with a significant optical component~\cite{Kavokin2007,Carusotto2013}. Their integer spin degree of freedom additionally allows for spin currents~\cite{Leyder2007,Kammann2012}. In fact, even without coupling to light, indirect excitons in double quantum wells also enable spin currents due to their long spin relaxation times~\cite{High2013}. A wide variety of important devices has been realized in recent years based on excitons and light-matter interactions, such as excitonic transistors~\cite{High2008,Grosso2009} with optical coupling and control~\cite{Andreakou2014}, and a range of polaritonic optical switches/transistors~\cite{Adrados2011,DeGiorgi2012,Anton2012,Cerna2013} or cascadable devices~\cite{Ballarini2013}. Exciton and polariton systems are both strongly influenced by disorder, however, which leads to resonant Rayleigh scattering and to the reduction of signals carried by ballistic particle propagation. Topology holds the promise to remedy the issues of disorder and backscattering, as was envisioned several years ago for excitons at the surface of 3D topological insulators~\cite{Seradjeh2009} and in coupled quantum wells close to ferromagnetic insulating films~\cite{Hao2010}, and more recently for exciton condensates in bilayer HgTe~\cite{Budich2014} and InAs/GaSb~\cite{Pikulin2014} quantum wells.

\begin{figure}[tbp]
\centering
\includegraphics[width=\linewidth]{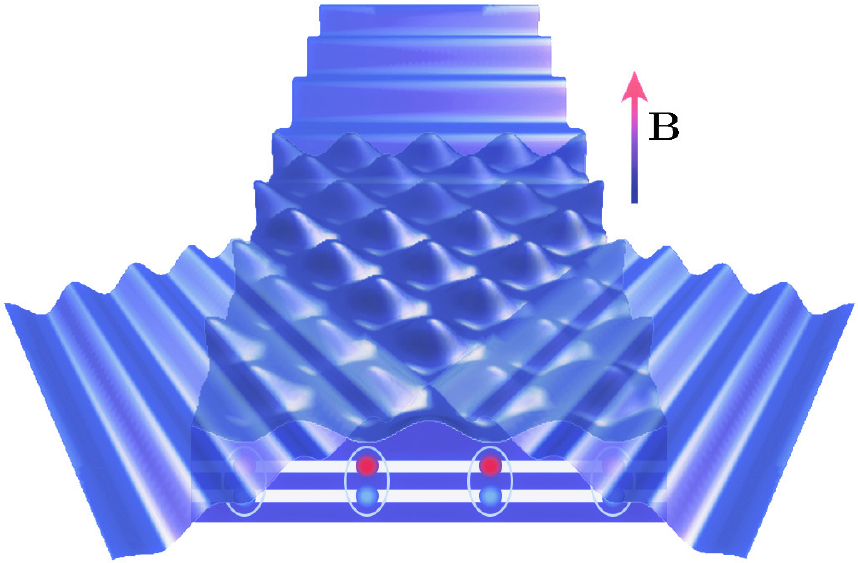}
\caption{(Color online) Schematic view of a typical system supporting topological polaritons or excitons: Surface acoustic waves modulate the thickness of quantum wells and interfere to generate a triangular lattice potential for excitons in the plane. Whether multiple quantum wells are coupled together, as in the depicted system of indirect excitons, or whether excitons are strongly mixed with light inside a microcavity, combining the periodic potential with an applied Zeeman field $\mathbf{B}$ leads to topologically non-trivial bands.}
\label{fig:system}
\end{figure}

In this Letter, we demonstrate theoretically that topological excitons and polaritons can be realized using surprisingly simple ingredients. We discuss the generic ingredients required to reveal this topological behavior, and demonstrate that our scheme can be realized in two very common types of systems: (i) exciton-polaritons in semiconductor microcavities, and (ii) indirect excitons in coupled quantum wells. In scenario (i), one or more quantum wells are typically placed at the optical antinodes of a microcavity so that excitons and cavity photons are strongly coupled. The resulting exciton-polaritons exhibit two spin states coupled by the transverse-electric-transverse-magnetic (TE-TM) splitting typical of microcavities~\cite{Panzarini1999}. This splitting arises mainly from the polarization-dependent reflectivity of distributed Bragg mirrors (leading to polarization-dependent energies of the optical modes), supplemented by a weaker but complementary polarization splitting of (direct) excitons stemming from the long-range exchange interaction between electrons and holes~\cite{Maialle1993}. In scenario (ii), we consider a pair of coupled quantum wells in close enough proximity for long-lived indirect excitons to form from electrons and holes in different layers. Four different spin states (corresponding to two dark and bright excitons) are present in that case, with Dresselhaus-type spin-orbit coupling~\cite{Matuszewski2012}.

In both scenarios, a magnetic field is required to break time-reversal symmetry. The natural sensitivity of excitons to applied magnetic fields circumvents the need for materials with large optical gyrotopic permeability and allows to operate at optical frequencies. A periodic exciton or polariton potential is also required to open a global (topological) gap, as discussed in Ref.~\cite{Karzig2014}. Such potential modulations can be implemented by applying surface acoustic waves~\cite{Rudolph2007,CerdaMendez2010,CerdaMendez2013}, as illustrated in Fig.~\ref{fig:system}, or by using permanent triangular/hexagonal lattices~\cite{Kim2013,Jacqmin2013}.


The simplification at the root of this paper stems from the linear-to-circular polarization conversion naturally present in garden variety systems of excitons and exciton-polaritons~\cite{Kavokin2005,Leyder2007,High2013}. Topological polaritons can be created by a ``winding coupling'' of topologically trivial exciton and photon bands~\cite{Karzig2014}. While an engineered winding coupling was considered in the original proposal of Ref.~\cite{Karzig2014}, here we exploit the fact that the TE-TM splitting of ordinary polariton bands naturally provides such a winding. In fact, a simple rotation from the momentum-dependent basis of TE-TM polarizations to a basis of circularly polarized states $(\psi_+, \psi_-)$ yields a coupling of the form
\begin{equation}
\mathcal{H}_\mathrm{TE-TM}\left(\begin{array}{c}\psi_+\\ \psi_-\end{array}\right)=\Delta k^2\left(\begin{array}{cc}0&e^{-2i\phi(\mathbf{k})}\\e^{2i\phi(\mathbf{k})}&0\end{array}\right)\left(\begin{array}{c}\psi_+\\ \psi_-\end{array}\right),
\label{eq:TETM_Hamiltonian}
\end{equation}
with a coupling strength proportional to the TE-TM splitting $\Delta$~\cite{Kavokin2005}. As desired, the coupling between modes with opposite polarizations winds (twice) in terms of the polar angle $\phi(\mathbf{k})$ associated with the in-plane wavevector $\mathbf{k}$. This winding coupling is readily accessible in experiments, and is well known for its crucial role in the optical spin Hall effect~\cite{Leyder2007,Kammann2012} and spin-to-angular momentum conversion~\cite{Manni2011}.

\begin{figure}
\includegraphics[width=\linewidth]{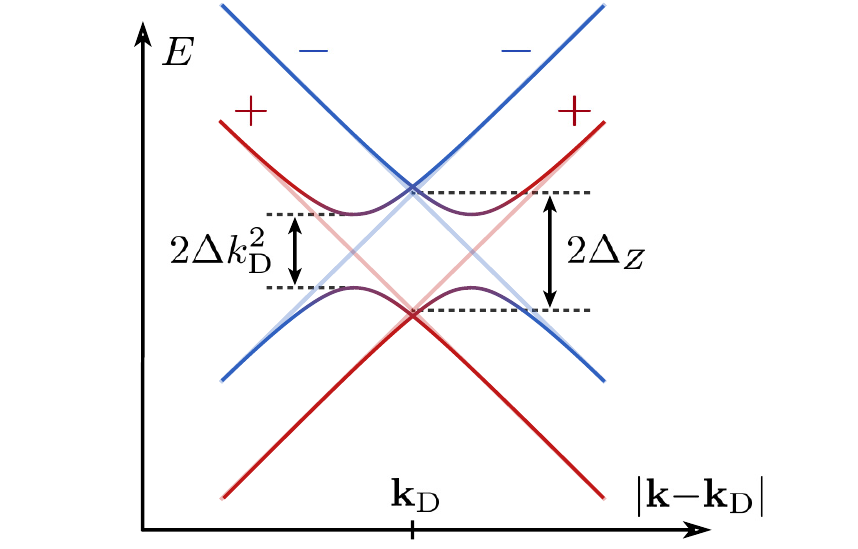}
\caption{Opening up a topological gap: Cross-sectional view of the typical band structure ``cut'' through a Dirac point located at $\mathbf{k}_\mathrm{D}$. A finite Zeeman field splits the bands corresponding to $+$ and $-$ circular polarizations into two Dirac cones (shown in faint red and blue) by $2\Delta_Z$, leading to a well-defined ring of resonance. The TE-TM (winding) coupling~\eqref{eq:TETM_Hamiltonian} of strength $\Delta k_\mathbf{D}^2$ then opens a topological gap, resulting in hybridized bands (solid lines) with Chern number $\pm 2$.}
\label{fig:tetmtopology}
\end{figure}

When two energy bands corresponding to distinct circular polarizations cross, an avoided-crossing gap opens up due to the winding coupling~\eqref{eq:TETM_Hamiltonian}, and the resulting hybridized bands exhibit a non-trivial topology. This can be understood from the fact that the spinor $(\psi_+, \psi_-)$ describing the two bands (with spin up/down or $+/-$ polarization) fully wraps the unit sphere when $\mathbf{k}$ runs over all momenta. Increasing $|\mathbf{k}|$ through the resonance yields a flip of this spinor, while the winding coupling $e^{-2i\phi(\mathbf{k})}$ leads to an azimuthal twist which completes the (double) wrapping of the unit sphere. The crossing between $+$ and $-$ bands required for such topological behavior is most easily obtained at isolated Dirac points of the spectrum generated by applying a triangular (or hexagonal) periodic potential~\cite{Kexin2014}. An applied Zeeman field then splits the $+$ and $-$ Dirac cones by $2\Delta_Z$, and a topological gap is opened along the resulting ring of resonance as a result of the TE-TM splitting (see Fig.~\ref{fig:tetmtopology}). Note that the Zeeman field provides the time-reversal symmetry breaking necessary for the appearance of quantum-Hall-like edge modes, as we demonstrate below. The periodic exciton potential, on the other hand, is required to open a global gap in the polariton or exciton spectrum, as detailed in Ref.~\cite{Karzig2014}.

To describe potentials of arbitrary strengths, below we derive the exact form of the energy spectrum expected for experimentally relevant potential profiles (rather than using an effective tight-binding description). For simplicity, we consider a sinusoidal potential that can be generated, e.g., by interfering surface acoustic waves as illustrated in Fig.~\ref{fig:system}, namely,
\begin{align}
V(x,y)&=V_0\left[\cos\left(\frac{4\pi}{\sqrt{3}a}y\right)+\cos\left(\frac{2\pi x}{a}+\frac{2\pi y}{\sqrt{3}a}\right)\right.\notag\\
&\hspace{10mm}\left.+\cos\left(\frac{2\pi x}{a}-\frac{2\pi y}{\sqrt{3}a}\right)\right],
\label{eq:CosPotential}
\end{align}
where $V_0$ is the amplitude of the potential and $a$ the lattice constant. Below we consider a ribbon-type geometry where the system is periodic in the $x$-direction and finite in the $y$-direction (with Dirichlet boundary conditions). The Sch\"odinger equation for the spinor polariton wavefunction reads
\begin{align}
&\left[-\frac{\hbar^2}{2m_\mathrm{eff}}\left(\frac{\partial^2}{\partial x^2}+\frac{\partial^2}{\partial y^2}\right)+V(x,y)+\sigma\Delta_Z-\varepsilon\right]\psi_\sigma(x,y)\notag\\
&\hspace{5mm}+\Delta\left(-\frac{\partial^2}{\partial x^2}+2i\sigma\frac{\partial^2}{\partial x\partial y}+\frac{\partial^2}{\partial y^2}\right)\psi_{-\sigma}(x,y)=0,
\label{eq:schrodingerEqPolaritons}
\end{align}
where $m_\mathrm{eff}$ is the polariton effective mass, $\sigma = \pm 1$ distinguishes the two circular polarizations, $\varepsilon$ is an energy eigenvalue, $\Delta$ is the strength of TE-TM splitting, and $2\Delta_Z$ is the Zeeman splitting caused by the applied magnetic field (in Faraday geometry). We remark that the latter acts on the magnetic moment of the excitonic component of the polaritons. More specifically, the splitting $\Delta_Z$ is related to the strength of the applied magnetic field $B$ via the electron and hole g-factors, i.e., $\Delta_Z=\frac{1}{2}(g_e-g_h)\mu_BB$ ($\mu_B$ being the Bohr magneton).

Translation symmetry in the $x$-direction makes it convenient to express the solutions of Eq.~\eqref{eq:schrodingerEqPolaritons} in Bloch form, i.e., $\psi_\sigma(x,t)=e^{ik_xx}u_\sigma(x,y)$ with $u_\sigma(x,y)$ periodic in $x$. The periodicity of $u_\sigma(x,y)$ and $V(x,y)$ allows to expand the latter as Fourier sums. Substitution into the Schr\"odinger equation~\eqref{eq:schrodingerEqPolaritons} then leads to an eigenvalue problem that can readily be solved numerically to obtain the energy spectrum $\varepsilon(k_x)$.

What values can be achieved in typical experiments for the various parameters introduced above? Zeeman splittings of up to $0.2$meV have been measured for exciton-polaritons in semiconductor microcavities under a magnetic field of $5$T~\cite{Walker2011}. An effective Zeeman splitting of up to $1$meV can also be optically induced by generating a large spin imbalance~\cite{Martin2002}, although smaller values are sufficient for our purposes, as we demonstrate below. So far exciton potentials with an amplitude of $0.18$meV have been reported in semiconductor microcavities using surface acoustic waves~\cite{CerdaMendez2010}. Higher values can however be envisioned given that amplitudes of up to $2$meV were reported in bare quantum wells~\cite{Rudolph2007}. Larger effective potential amplitudes can also be expected for permanent polariton potentials realized by patterning composite materials~\cite{Kim2013,Jacqmin2013}. Typical values for the TE-TM splitting are around $\Delta=0.05$meV$\mu$m$^2$~\cite{Kammann2012}.

\begin{figure}[tbp]
\centering
\includegraphics[width=\linewidth]{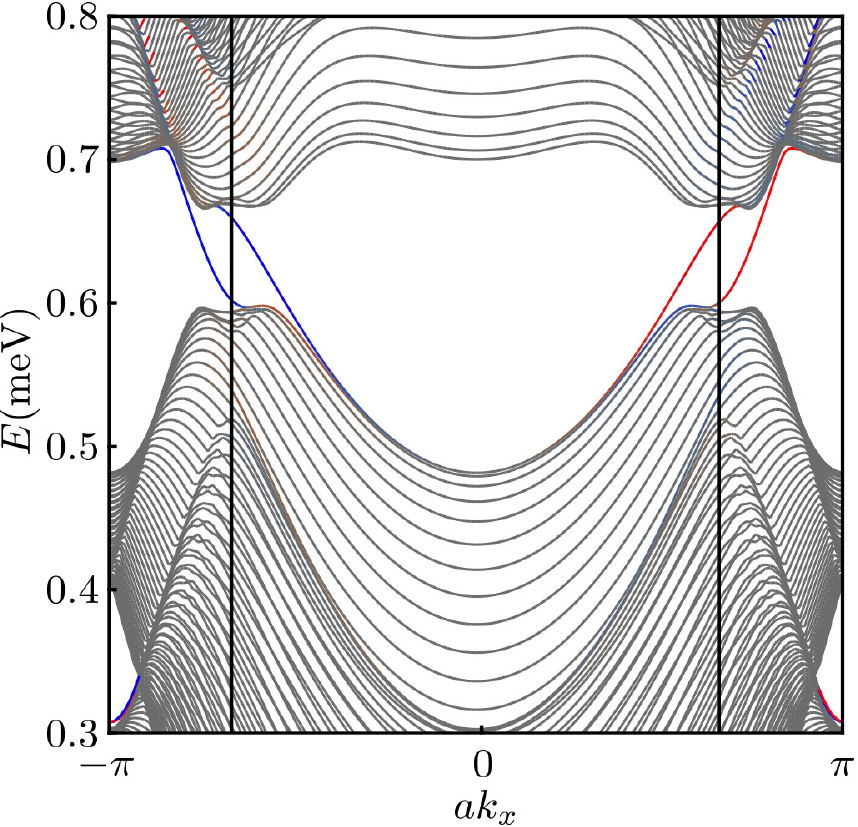}
\caption{(Color online) Typical energy dispersion of topological exciton-polaritons in a triangular lattice potential of the form~\eqref{eq:CosPotential} with periodic boundary conditions in the $x$-direction and vanishing (Dirichlet) boundaries conditions in the $y$-direction. Eigenstates are color-coded according to their proximity to edges --- red and blue corresponding to lower and upper edges, respectively --- with bulk states shown in grey. The vertical lines indicate the positions of Dirac points at $ak_x=\pm2\pi/3$. Parameters: $\Delta=0.05$meV$\mu$m$^2$, $2\Delta_Z=0.1$meV, $V_0=0.6$meV, $a=3\mu$m, and $m_\mathrm{eff}=7.5\times10^{-5}m_0$~\cite{Kammann2012}, where $m_0$ is the free electron mass.}
\label{fig:SpinorPolariton}
\end{figure}

We present in Fig.~\ref{fig:SpinorPolariton} the form of a typical topological polariton dispersion obtained for conservative experimentally available parameters. Here a typical TE-TM splitting of $\Delta=0.05$meV$\mu$m$^2$ combined with a Zeeman splitting of $2\Delta_Z=0.1$meV and a periodic potential of amplitude $V_0=0.6$meV induces a topological gap of the order of $0.7$meV, which is well within resolvable range with typical polariton linewidths of the order of tens of $\mu$eV~\cite{Carusotto2013}. Larger gaps can be achieved, e.g., at higher fields. As anticipated above, the gap is bridged by two pairs of chiral edge states localized at opposite edges. In accordance with bulk-edge correspondence arguments (see, e.g., Ref.~\cite{Hasan2010}), the lower and upper bands are topologically non-trivial with Chern numbers $+2$ and $-2$, respectively.

\begin{figure}[tbp]
\centering
\includegraphics[width=\linewidth]{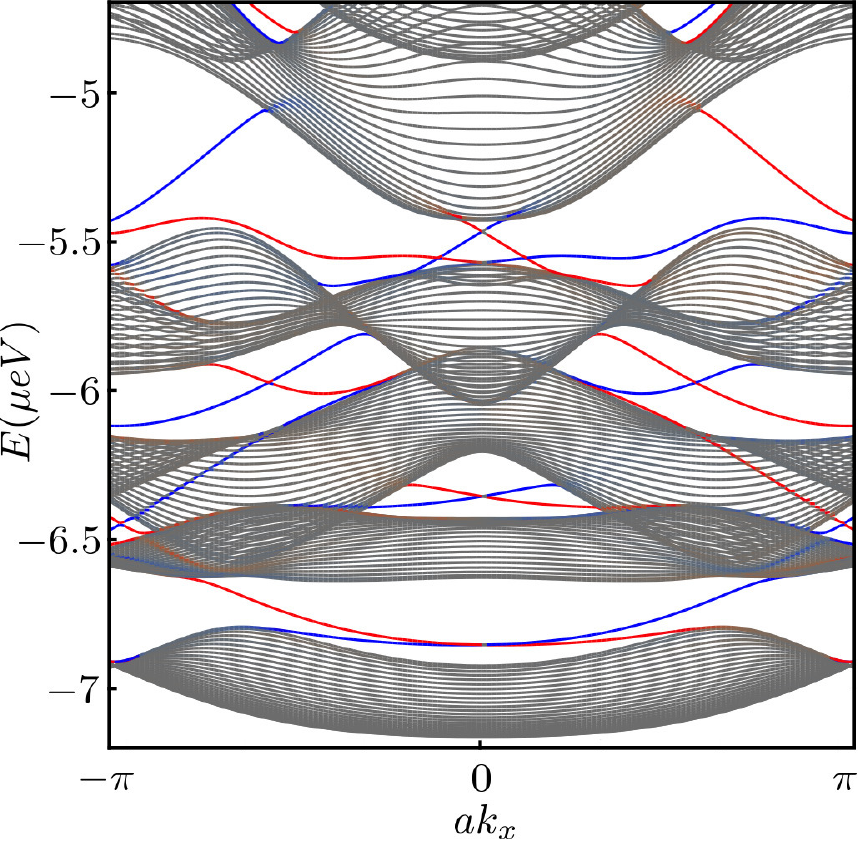}
\caption{(Color online) Typical energy dispersion of topological indirect excitons with similar color-coding as in Fig.~\ref{fig:SpinorPolariton}. Spin-orbit coupling parameters, g-factors and effective masses were taken from Ref.~\cite{Kavokin2013}: $\beta_e=2.7\mu$eV$\mu$m, $\beta_h=0.92\mu$eV$\mu$m, $g_e=0.01$, $g_h=-8.5\times10^{-3}$, $m_e=0.07m_0$, and $m_h=0.16m_0$, where $m_0$ is the free electron mass. A magnetic field $B=2$T was assumed, along with a periodic triangular lattice potential of amplitude $V_0=5\mu$eV with lattice constant $a=1.2\mu$m.}
\label{fig:IndirectExciton}
\end{figure}

An even more appealing platform for realizing chiral edge states in exciton-based systems is provided by {\it indirect excitons}. The latter are typically formed in structures of coupled quantum wells where electrons and holes are confined in separate wells. More importantly, they are perhaps best known for their long radiative lifetime stemming from the reduced overlap between electron and hole wavefunctions, which allows for the formation of condensates~\cite{High2012}. Indirect excitons are also appreciated as having a rich four-component spin degree of freedom~\cite{Rubo2011} due to the co-existence of bright excitons (with $J_z=\pm1$ spin projections normal to the plane) and dark excitons (with $J_z=\pm2$) at similar energies, as well as a diverse spin dynamics due to spin-orbit interactions~\cite{Matuszewski2012,Kyriienko2012}. Of particular interest here is the Dresselhaus spin-orbit coupling arising from the intrinsic crystal asymmetry of zinc-blende (e.g., GaAs) crystals, which was also invoked in Ref.~\cite{Hao2010}. In a basis of exciton spinor wavefunctions $(\psi_{+1},\psi_{-1},\psi_{+2},\psi_{-2})$, this coupling can be described by a Hamiltonian of the form
\begin{equation}
\mathcal{H}_\mathrm{ex}=\left(\begin{array}{cccc}\Delta_Z&0&\beta_e k_e e^{-i\phi}&\beta_h k_h e^{-i\phi}\\0&-\Delta_Z&\beta_h k_h e^{i\phi}&\beta_e k_e e^{i\phi}\\ \beta_e k_e e^{i\phi}&\beta_h k_h e^{-i\phi}&-\Delta_Z'&0\\ \beta_h k_h e^{i\phi}&\beta_e k_e e^{-i\phi}&0&\Delta_Z'\end{array}\right),
\label{eq:ExcitonSpinCoupling}
\end{equation}
where $\beta_e$ and $\beta_h$ are Dresselhaus constants for electrons and holes, respectively~\cite{Kavokin2013}. The wavevectors associated with electrons and holes with effective masses $m_e$ and $m_h$ are related to those of excitons via $\mathbf{k}_e=\tfrac{m_e}{m_e+m_h}\mathbf{k}$ and $\mathbf{k}_h=\tfrac{m_h}{m_e+m_h}\mathbf{k}$, respectively. Here the Zeeman splitting is different for bright and dark excitons due to the distinct spin orientations of their electron and hole components: While $\Delta_Z=\tfrac{1}{2}(g_e-g_h)\mu_BB$ for bright excitons as above, $\Delta_Z'=-\tfrac{1}{2}(g_e+g_h)\mu_BB$ for dark excitons. Note that we have neglected the Rashba spin-orbit coupling~\cite{Kyriienko2012,Kavokin2013}, for simplicity, since the latter is only significant in the presence of a bulk quantum-well asymmetry or under a large electrical bias.

Using the exciton-spin coupling Hamiltonian~\eqref{eq:ExcitonSpinCoupling}, we now examine a system of indirect excitons with the same ribbon geometry and periodic exciton potential as above (which can similarly be realized, e.g., using surface acoustic waves~\cite{Rudolph2007}). The method introduced in Eq.~\eqref{eq:schrodingerEqPolaritons} readily generalizes to the present scenario: Here the analog of Eq.~\eqref{eq:schrodingerEqPolaritons} involves four spin components $+1, -1$ and $+2, -2$ corresponding to bright and dark excitons, respectively, the relevant effective mass is the exciton mass $m_\mathrm{ex}=m_e+m_h$, and the spin-orbit coupling and magnetic field are taken into account as described by Eq.~\eqref{eq:ExcitonSpinCoupling}. In the absence of any periodic potential, the low-lying branch of the exciton dispersion has a minimum at $k\neq0$, which is a well-known consequence of the linear growth of the Dresselhaus coupling strength with $k$.

The four-band nature of the system combined with the appearance of single windings $e^{i\phi}$ and $e^{-i\phi}$ in the coupling Hamiltonian~\eqref{eq:ExcitonSpinCoupling} generically results in rich features upon introduction of a periodic potential (see Fig.~\ref{fig:IndirectExciton}). Remarkably, the system enables multiple bandgaps with different numbers of topologically protected chiral edge states. Bandgaps are either (i) topologically trivial, (ii) topological with a single pair of counter-propagating chiral edge states, or (iii) topological with two such pairs. We present in Fig.~\ref{fig:IndirectExciton} the typical indirect-exciton dispersion obtained using experimentally available parameters (taken from Ref.~\cite{Kavokin2013}). Two low-energy gaps are shown, exemplifying both cases (ii) and (iii). In practice, parameters can be chosen so as to optimize the size of a particular gap: Under a magnetic field $B=2$T with spin-orbit coupling and indirect-exciton parameters from Ref.~\cite{Kavokin2013}, the lower energy gap (of type (ii)) can reach about $0.3\mu$eV with a periodic potential of amplitude $V_0=6\mu$eV with lattice constant $a=1.1\mu$m, while a gap of type (iii) of about $1.3\mu$eV can be obtained by choosing $V_0=10\mu$eV and $a=0.75\mu$m. Note that the maximum achievable gap generically increases linearly with the applied magnetic field $B$. Since indirect excitons are typically much longer-lived than polaritons~\cite{High2012}, topological gaps of the order of $1\mu$eV are in principle well within resolvable range.

We remark that indirect excitons may ultimately offer greater potential for topological devices as compared to polaritons. The main limitation for the ballistic propagation of polaritons is their short radiative lifetime (in the range of tens of picoseconds~\cite{Carusotto2013}). While topologically protected states of polaritons would be useful for reducing errors in photonic devices, amplification would still be required to maintain propagation~\cite{Wertz2012}. Indirect excitons, on the other hand, exhibit relatively long lifetimes (up to milliseconds~\cite{High2012}), leaving disorder as a more significant factor in the control of ballistic propagation.


In this manuscript, we have demonstrated that topological polaritons and excitons can be realized in garden-variety single and double quantum wells made of ordinary materials such as GaAs. The key ingredient to reveal their topological behavior is a triangular/hexagonal lattice potential which can be realized, for example, using surface acoustic waves. The resulting Dirac points in the dispersion split under the combined effect of an applied magnetic field (in Faraday configuration) and spin-orbit coupling, giving rise to a topological bandgap bridged by topologically protected chiral edge states. Here we make use of the TE-TM splitting and Dresselhaus-type spin-orbit coupling naturally present in microcavities and in systems of indirect excitons, respectively. Conservative experimental values for these couplings give rise to topological gaps that are larger than the typical polariton/exciton linewidth with readily available magnetic fields (well below $5$T) and exciton potential amplitudes (below $1$meV). We expect the ability to engineer edge states protected from backscattering using these simple available ingredients to play an important role in the development of a variety of exciton-based information processing devices. The study of nonlinear interactions and their effect on such topological states is an important direction for future work.

While we were finalizing our manuscript, a related proposal for polaritons in tight-binding lattices made from micropillar arrays was posted on the arXiv~\cite{Nalitov2014}. Our study demonstrates that topological states are accessible in a range of exciton-based systems with experimentally available parameters, and highlights indirect excitons as a more promising platform for topological devices.


This work was funded by the Institute for Quantum Information and Matter, an NSF Physics Frontiers Center with
support of the Gordon and Betty Moore Foundation through Grant GBMF1250. G. R. acknowledges the Packard Foundation and NSF under DMR-1410435 for their generous support. Financial support from the Swiss National Science Foundation (SNSF) is also gratefully acknowledged.

* These authors contributed equally.


\end{document}